\documentclass[pre,amssymb,noshowpacs]{revtex4}
\usepackage{graphicx}
\usepackage{dcolumn}
\usepackage{bm}
\begin{document}
\title{Demixing in athermal mixtures of colloids and excluded-volume
polymers from a density functional theory}
\author{Pawe{\l} Bryk}
\affiliation{Department for the Modeling of Physico-Chemical Processes,
Maria Curie-Sk{\l}odowska University, 20-031 Lublin, Poland}
\email{pawel@paco.umcs.lublin.pl}
\date{\today}
\begin{abstract}
We study the structure and interfacial properties of model athermal mixtures of colloids and excluded volume polymers.
The colloid particles are modeled as hard spheres whereas the polymer coils are modeled as chains
formed from tangentially bonded hard spheres. 
Within the framework of the nonlocal density functional theory we study
the influence of the chain length on the surface tension and the interfacial width.
We find that the interfacial tension of the colloid-interacting polymer mixtures increases
with the chain length and is significantly smaller than that of the ideal polymers. 
For certain parameters we find oscillations on the colloid-rich parts of the density profiles of both colloids
and polymers with the oscillation period of the order of the colloid diameter. The interfacial width is 
few colloid diameters wide and also increases with the chain length. We find the interfacial width
for the end segments to be larger than that for the middle segments and this effect is more pronounced
for longer chains. 
\end{abstract}
\maketitle
\section{Introduction}
Suspensions of sterically stabilized colloids
and nonadsorbing polymers often exhibit phase separation \cite{Tuinier03,Poon02}.
Since the experimental parameters can be tailored to match
the desired properties, such systems have become an important tool
in investigating various theoretical concepts.
Recent experiments on well-characterized colloid-polymer mixtures
have involved studies of the wetting transition \cite{wet1,wet2},
real space observation of the thermal capillary waves \cite{Aarts04a},
the fluid-fluid interfacial tension \cite{Aarts03,Hoog99a,Chen00,Chen01}, and
the interfacial width \cite{Hoog99b}.

One of the first theoretical models giving an insight into this phenomenon 
is the Asakura-Oosawa-Vrij (AOV) model of colloid-polymer mixtures
\cite{Asakura54,Vrij76}, whereby the polymer-polymer interactions are ideal 
but the polymer-colloid and colloid-colloid interactions are of the hard-sphere
type. Demixing transition into the colloid-rich and the polymer-rich
phases may be explained by invoking the concept of attractive depletion interactions
that arise due to a tendency of the system to reduce the volume excluded to 
the centers of polymer coils \cite{Gast83,Lekkerkerker92}.
Brader and Evans \cite{Brader00} have investigated the interfacial properties of the AOV model within
the square gradient approximation. Their results were in a reasonable agreement with the experimental data.
Subsequently Schmidt {\it et al}. \cite{Schmidt00} have proposed  an extension of Rosenfeld's successful fundamental
measure theory \cite{Rosenfeld89} to the AOV model. It has turned out that this functional yields
the interfacial tension in a good agreement with the experiment and predicts a sequence of layering 
transitions in the partial wetting regime prior to a transition to complete wetting  \cite{Brader02a}.
These findings have been confirmed by Monte Carlo simulations \cite{Dijkstra02} and
recent experiments \cite{wet1,wet2}.

The Asakura-Oosawa-Vrij model was originally introduced for systems, where the polymer coils
are smaller than the colloid dimensions and the polymer-polymer interactions
are neglected. The latter approximation is reasonable if the second virial coefficient
of the monomer-monomer interaction vanishes what corresponds to the theta-point conditions.
The influence of the polymer-polymer interactions has been studied first by Warren {\it et al}.
\cite{Warren95} who used the perturbation theory around the AOV reference model.
Fuchs and Schweizer \cite{Fuchs00,Fuchs01,Fuchs02} have noted
that the polymer-polymer interactions lead to an upward shift of the spinodal.
These findings have been confirmed by the results of the modified free volume theory
\cite{Aarts02}, the geometry-based density functional approach \cite{Schmidt03}
and sophisticated computer simulation techniques \cite{Bolhuis02}.
Very recently the interfacial behavior of the interacting polymer-hard sphere colloid
mixtures within the square gradient approximation has been investigated \cite{Aarts04b,Moncho03b}.
An important ingredient of the square gradient approximation is an expression
for the direct correlation function \cite{Evans92}. This is usually tackled by employing
the mean-spherical approximation, so that only an expression for the interaction potential
is needed. Both groups have adopted different approaches.
To obtain the interaction potential between two colloidal spheres mediated by excluded volume
polymers Aarts {\it et al}. \cite{Aarts04b} have used the generalized Gibbs adsorption equation approach
\cite{Tuinier02}, whereas Moncho-Jorda {\it et al}. \cite{Moncho03b} have used
a  depletion potential obtained from direct computer simulations \cite{Louis02}.
Both groups have found that the interfacial tension is lower for interacting 
polymers than for the AOV model.

In a recent work Paricaud {\it et al}. \cite{Paricaud03}
have studied the phase behavior in model athermal mixtures
of colloids and freely jointed tangentially bonded hard sphere polymers
by means of Wertheim's first order thermodynamic perturbation theory (TPT1) \cite{Wertheim87}.
They have found that if the colloid diameter is much larger than the polymer segment diameter
the system may undergo a demixing transition. This theory offers several advantages over previously
mentioned approaches. As a fully microscopic theory it is free of the coarse-graining, which is a powerful concept
on its own \cite{Louis00}, but may be difficult to implement, for instance, close to geometrically structured substrates
\cite{Bryk03,Li03}. The two-body potentials resulting from the coarse-graining techniques may
be insufficient to correctly describe the underlying complex system
and the higher-order terms may have to be taken into account \cite{Moncho03a}.
Also, within the TPT1 theory it is straightforward to incorporate the attractive interactions.
However, the Wertheim TPT1 description is not free of its own caveats.
For example, an extension of the theory to the solid phase while in principle possible \cite{Vega01},
it is difficult to carry out for the colloid-polymer mixtures. Within the TPT1 theory
the second virial coefficient scales linearly with the chain length $M$ \cite{Boublik89,Boublik90}
instead of $\sim M^{3\nu}$, where $\nu$ is the Flory exponent. The consequence
is that within the TPT1 treatment the dilute regime is not described accurately.
Although it is possible to improve this deficiency \cite{Vega00}, it is not clear how
this approach could be extended to the colloid-polymer mixtures.
Despite these issues, the approach of Paricaud {\it et al}. \cite{Paricaud03}
seems to be quite interesting.

In the present work we apply the microscopic density functional theory of Yu and Wu
\cite{Yu02} to study the interfacial properties of the hard sphere colloids and
excluded volume polymers modeled as freely jointed tangentially bonded hard spheres.    
In the bulk limit this functional yields an equation of state identical
with that of Ref.~\cite{Paricaud03}. 
The applied functional is nonlocal, therefore it should correctly capture the structure of the interface
including possible oscillatory behavior of the density profiles. Also, the Yu and Wu theory
treats the polymer chains and colloid spheres on equal footing, therefore the many-body effects are
automatically included.

\section{Theory}
\label{sec:theory}
In this work we consider mixtures of colloids (species $C$)
and polymers (species $P$). The colloidal particles are represented as
hard spheres of diameter $\sigma_C$ whereas the polymer coils are
modeled as chains built of $M$ tangentially bonded hard-sphere segments of 
diameter $\sigma_{PS}$. 
We assume that there are no torsional or bending potentials
imposed on the polymer segments, i.e. the monomers
are freely jointed and any arbitrary polymer configuration free of the 
intermolecular and intramolecular overlaps is allowed.
By introducing the total bonding potential $V_{b}({\bf R})$ as a sum
of bonding potentials $v_b$ between the monomers
$V_{b}({\bf R})=\sum_{i=1}^{M-1}v_b(|{\bf r}_{i+1}-{\bf r}_{i}|)$
the intramolecular interactions for the freely-jointed tangential hard spheres
can be conveniently written as
\begin{equation}\label{eq:1}
\exp [-\beta V_{b}({\bf R})]=
\prod_{i=1}^{M-1}\frac{\delta(|{\bf r}_{i+1}-{\bf r}_{i}|-\sigma_{PS})}{4\pi\sigma_{PS}^{2}}.
\end{equation}
In the above ${\bf R}\equiv ({\bf r}_{1}, {\bf r}_{2}, \cdots, {\bf r}_{M})$ denotes
a set of monomer coordinates.

As in every density functional theory the
grand potential of the system is assumed to be a functional of the local densities of polymers
$\rho_{P}({\bf R})$ and colloids $\rho_{C}({\bf r})$
\begin{eqnarray}\label{eq:2}
\lefteqn{\Omega[\rho_P({\bf R}),\rho_{C}({\bf r})]=
F[\rho_P({\bf R}),\rho_{C}({\bf r})]+}\nonumber\\
& &\int\!\!d{\bf R}\rho_{P}({\bf R})(V_{ext}^{(P)}({\bf R})-\mu_{P})\nonumber\\
&+&\int\!\!d{\bf r}\rho_{C}({\bf r})
(V_{ext}^{(C)}({\bf r})-\mu_{C})\;,
\end{eqnarray}
where $V_{ext}^{(P)}({\bf R})$, $\mu_P$,  $V_{ext}^{(C)}({\bf r})$
and $\mu_C$ are the external and the chemical potentials for
polymers and colloids, respectively. 
The free energy of the system $F$ is a sum of
the ideal and excess contributions, $F=F_{id}+F_{ex}$.
The ideal part of the free energy is known exactly 
\begin{eqnarray}\label{eq:3}
\lefteqn{\beta F_{id}[\rho_P({\bf R}),\rho_{C}({\bf r})]=
\beta\int\!\!d{\bf R}\rho_{P}({\bf R})V_{b}({\bf R})}\nonumber\\
&+&\int\!\!d{\bf R}\rho_{P}({\bf R})[\ln(\rho_{P}({\bf R}))-1]\nonumber\\
&+&\int\!\!d{\bf r} \rho_{C}({\bf r})
[\ln(\rho_{C}({\bf r}))-1]\;,
\end{eqnarray}
where the dependence on the irrelevant thermal wavelengths for 
polymers and colloids was suppressed.

In their study of bulk properties of athermal mixtures
of colloid and excluded-volume polymers 
Paricaud {\it et al.} \cite{Paricaud03} employed the first order thermodynamic
perturbation theory of Wertheim \cite{Wertheim87}. 
Within this approach the excess free 
energy of the polymer system can be treated as a sum of the excess
free energy of the reference system containing unbonded monomers
and a perturbation due to formation of the polymer coils. 
Yu and Wu \cite{Yu02} incorporated Wertheim's first-order perturbation theory 
into the framework of the fundamental measure theory (FMT) of Rosenfeld \cite{Rosenfeld89}
providing thus a convenient mean of description of inhomogeneous mixtures
of tangentially jointed hard-sphere chains.
The connection between TPT1 and FMT can be made by assuming that $F_{ex}$ 
is a functional of the local density of colloids and average segment densities \cite{Yu02}
$\rho_{PS}({\bf r})$ defined as
\begin{equation}\label{eq:4}
\rho_{PS}({\bf r})=\sum_{i=1}^{M}\rho_{PS,i}({\bf r})=\sum_{i=1}^{M}
\int\!\!d{\bf R}\delta({\bf r}-{\bf r}_i)\rho_{P}({\bf R})\;,
\end{equation}
where $\rho_{PS,i}({\bf r})$ is the local density of the polymer segment $i$.
The convolutions of the local densities $\rho_{j}({\bf r})$, $j=PS, C$
with a suitable set of weight functions $w_{\alpha}^{(j)}({\bf r})$, $\alpha=3,2,1,0,V2,V1$
yield the weighted densities $n_{\alpha}=n_{\alpha}^{(PS)}+n_{\alpha}^{(C)}$
\begin{equation}\label{eq:5}
n_{\alpha}^{(j)}({\bf r})=\int\!\!d{\bf r}' \rho_{j}({\bf r}')
w_{\alpha}^{(j)}({\bf r}-{\bf r}')\;.
\end{equation}
The weight functions $w_{\alpha}^{(j)}({\bf r})$ are related to the geometric properties
of colloidal particles and monomer segments \cite{Rosenfeld89}
\begin{eqnarray}\label{eq:6}
w^{(j)}_{3}({\bf r})=\Theta(\frac{\sigma_j}{2}-|{\bf r}|)\;&,&\;
w_{2}^{(j)}({\bf r})=\delta (\frac{\sigma_j}{2}-|{\bf r}|)\;,\\
{\bm w}_{V2}^{(j)}({\bf r})=\frac{{\bf r}}{|{\bf r}|}\delta (\frac{\sigma_j}{2}-|{\bf r}|)\;
&,& w_{1}^{(j)}({\bf r})=\frac{w_{2}^{(j)}({\bf r})}{2\pi\sigma_j}\;,\\
w_{0}^{(j)}({\bf r})=\frac{w_{2}^{(j)}({\bf r})}{\pi\sigma_j^2}\;
&,&{\bm w}_{V1}^{(j)}({\bf r})=\frac{{\bm w}_{V2}^{(j)}({\bf r})}{2\pi\sigma_j}\;.
\end{eqnarray}

In the FMT approach the excess free energy is obtained as a volume integral
over the free energy density $F_{ex}=\int d{\bf r}\Phi$ expressed as
a simple function of the weighted densities. 
Following Yu and Wu \cite{Yu02} we assume that $\Phi=\Phi_{P}+\Phi_{HS}$,
where $\Phi_{HS}$ describes the reference mixture of hard spheres of 
diameters $\sigma_C$ and $\sigma_{PS}$, 
while the excess free energy density due to the chain connectivity $\Phi_{P}$ is 
an ``inhomogeneous counterpart'' of the perturbation term in TPT1. 

There are several expressions for the hard-sphere part $\Phi_{HS}$.
For the present problem we choose the elegant and inspiring White-Bear version of the FMT
\cite{Roth02b,Yu02b}
\begin{eqnarray}\label{eq:9}
\Phi_{HS}(\{n_{\alpha}\})=-n_0 \ln (1-n_{3})+
\frac{n_{1}n_{2}-{\bm n}_{V1}\cdot 
{\bm n}_{V2}}{1-n_{3}}&&\nonumber\\
+(n_2^3-3n_2{\bm n}_{V2}\cdot{\bm n}_{V2})\frac{n_{3}+
(1-n_3)^2\ln (1-n_3)} 
{36\pi (n_3)^2(1-n_{3})^{2}}&&\,.
\end{eqnarray}

The contribution $\Phi_{P}$ due to chain connectivity
is evaluated using Wertheim's first-order
perturbation theory \cite{Wertheim87,Yu02}
\begin{equation}  \label{eq:10}
\Phi_{P}(\{n_{\alpha}\};\{n_{\alpha}^{(PS)}\};\sigma_{PS})=\frac{1-M}{M}n_0^{(PS)}\zeta^{(PS)}\ln
[y_{HS}(\sigma_{PS}; \{n_{\alpha}\})]\;,
\end{equation}
where $\zeta^{(PS)}=1-\mathbf{n}_{V2}^{(PS)}\cdot \mathbf{n}_{V2}^{(PS)}/(n_2^{(PS)})^2$.
$y_{HS}$ is the Boublik-Mansoori-Carnahan-Starling-Leland \cite{Boublik70,Mansoori71} expression
for the contact value of the radial distribution function of mixtures
hard spheres
\begin{equation}  \label{eq:11}
y_{HS}(\sigma_{PS}; \{n_{\alpha}\})=\frac 1{1-n_3}+\frac{n_2\sigma_{PS}\zeta }{%
4(1-n_3)^2}+\frac{(n_2\sigma_{PS})^2\zeta }{72(1-n_3)^3}\;.
\end{equation}
with $\zeta =1-\mathbf{n}_{V2}\cdot \mathbf{n}_{V2}/(n_2)^2$.

At equilibrium the first functional derivative of the grand potential
with respect to the densities of colloids and polymers vanishes
\begin{equation}\label{eq:12}
\frac{\delta\Omega [\rho_P({\bf R}),\rho_{C}({\bf r})]}
{\delta \rho_P(\mathbf{R})}=
\frac{\delta\Omega [\rho_P({\bf R}),\rho_{C}({\bf r})]}
{\delta \rho_C({\bf r})}=0 \;,
\end{equation}
which leads to the following equation for the
average segment density profile
\begin{equation}\label{eq:13}
\rho_{PS}(\mathbf{r})=\exp (\beta \mu_P )\int d\mathbf{R}%
\sum_{i=1}^{M}\delta (\mathbf{r}-\mathbf{r}_i)\exp \left[ -\beta
V_b(\mathbf{R})-\beta \sum_{l=1}^{M}\lambda _l(\mathbf{r}%
_l)\right] \;,
\end{equation}
where $\lambda_l(\mathbf{r}_l)$ is
\begin{equation}\label{eq:14}
\lambda_l(\mathbf{r}_l)=\frac{\delta F_{ex}}{\delta \rho_{PS}(%
\mathbf{r}_l)}+v_{l}(\mathbf{r}_l)\;,
\end{equation}
with $v_{l}(\mathbf{r}_l)$ being an external potential acting on the $l$-th segment.
Equation (\ref{eq:13}) can be rewritten as
\begin{equation}  \label{eq:15}
\rho_{PS}({\bf r})=\exp(\beta \mu_P)\sum_{i=1}^{M}\exp [-\beta
\lambda_i({\bf r})]G_i({\bf r})G_{M+1-i}({\bf r})\;,
\end{equation}
where the propagator function $G_i({\bf r})$ is determined
from the recurrence relation
\begin{equation}  \label{eq:16}
G_i({\bf r})=\int d {\bf r}^{\prime }
\exp [-\beta \lambda_i({\bf r}^{\prime})]%
\frac{\delta (\sigma_{PS} -|{\bf r} -{\bf r}^{\prime }|)}{4\pi \sigma_{PS}^2 }%
G_{j-1}({\bf r}^{\prime })
\end{equation}
for $i=2,3,\dots ,M$ and with $G_1({\bf r})\equiv 1$.
Likewise Eq.~(\ref{eq:12}) leads to the following equation for the density profile
of colloids
\begin{equation}\label{eq:17}
\rho_{C}(\mathbf{r})=\exp\left[\beta \mu_C 
-\beta V_{ext}^{(C)}(\mathbf{r})-\beta\delta F_{ex}/\delta\rho_C({\bf r})\right] \;.
\end{equation}
Equations (\ref{eq:15}) and (\ref{eq:17}) can be solved numerically using the standard Picard iterative method.
\section{Results}
In homogeneous systems the thermodynamic properties can be calculated analytically. 
To this end we note that in the bulk the vector weighted densities vanish 
whereas the scalar weighted densities become proportional to the 
corresponding bulk densities. For the polymer weighted densities we
have $n_{\alpha}^{(PS)}=\xi_{\alpha}^{(P)} M\rho^{(b)}_{P}=\xi_{\alpha}^{(P)}\rho^{(b)}_{PS}$, 
with $\xi_3^{(P)}=\pi/6\,\sigma_{PS}^3$, $\xi_2^{(P)}=\pi\sigma_{PS}^2$,
$\xi_1^{(P)}=\sigma_{PS}/2$ and $\xi_0^{(P)}=1$. Similar, for the colloid weighted densities we have
$n_{\alpha}^{(C)}=\xi_{\alpha}^{(C)} \rho^{(b)}_{C}$. 
The total configurational free energy per unit volume, $\Phi_v$,
can be straightforwardly obtained by inserting the expressions for the weighted densities into
Eqs. [(\ref{eq:9})-(\ref{eq:11})], $\Phi_v=\Phi_{HS}+\Phi_{P}+
\beta^{-1}\rho^{(b)}_{P}[\ln(\rho^{(b)}_{P})-1]+\beta^{-1}\rho^{(b)}_{C}[\ln(\rho^{(b)}_{C})-1]$.
Finally, the pressure $P$, and the chemical potentials of both species are
determined from
\begin{equation}\label{eq:18}
\beta P=-\Phi_v+\sum_{j=P,C}\rho^{(b)}_j\frac{\partial\Phi_v}
{\partial \rho^{(b)}_j}\;,\;\;\beta\mu_j=\frac{\partial\Phi_v}{\partial\rho^{(b)}_j}\;.
\end{equation}

For bulk systems our theory is identical to the Wertheim TPT1 theory for athermal
mixtures of hard spheres and excluded-volume polymers.
Paricaud {\it et al.} \cite{Paricaud03} conjectured that TPT1 predicts a demixing transition
into colloid-rich (polymer-poor) and colloid-poor (polymer-rich) phases.
The coexisting equilibrium densities (binodals)  
were obtained from the condition of the equality of pressure and chemical potentials of
both species in the two demixed phases. The spinodal lines delimiting the
regions of the stability against fluctuations of density and composition
were calculated from 
$\det[\partial^2\Phi_v/\partial\rho^{(b)}_i\partial\rho^{(b)}_j]=0$, $i,j=P,C$.
The critical points were evaluated from \cite{Rowlinson59}
\begin{eqnarray}\label{eq:21}
s^3\frac{\partial^3\Phi_v}{\partial[\rho^{(b)}_P]^3}+3s^2
\frac{\partial^3\Phi_v}{\partial[\rho^{(b)}_P]^2\partial\rho^{(b)}_C}&+&
3s\frac{\partial^3\Phi_v}{\partial\rho^{(b)}_P\partial[\rho^{(b)}_C]^2}+
\frac{\partial^3\Phi_v}{\partial[\rho^{(b)}_C]^3}=0\,,\nonumber\\
s\equiv\frac{-\partial^2\Phi_v}{\partial\rho^{(b)}_P\partial\rho^{(b)}_C}&/&
\frac{\partial^2\Phi_v}{\partial[\rho^{(b)}_P]^2}\;.
\end{eqnarray}

\begin{figure}
\includegraphics[clip,width=6.5cm]{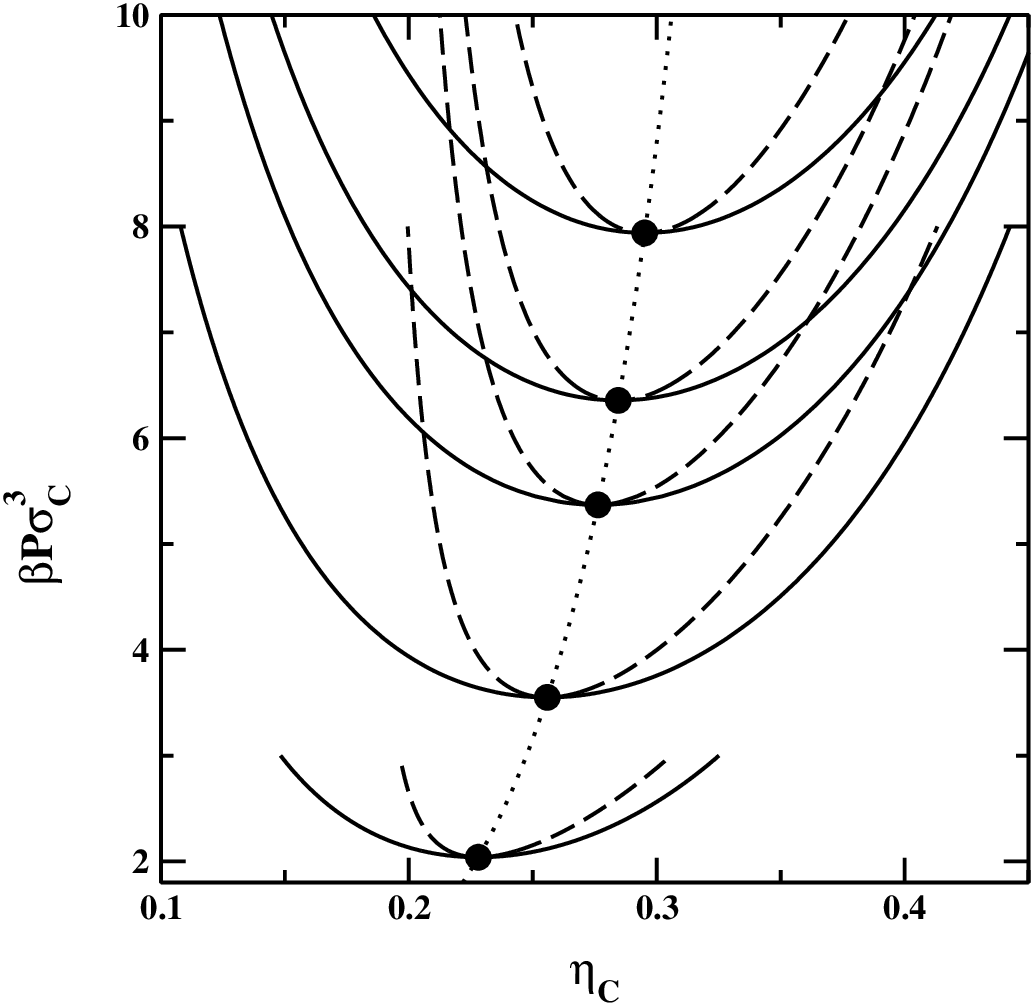}
\caption{\label{fig:1}
The binodals (solid lines), the spinodals (dashed lines) and the critical
points (black dots) for the demixing transition in colloid-excluded volume
polymer mixtures resulting from the TPT1 theory. The phase
diagrams are evaluated for the constant ratio of colloid diameter to polymer segment
diameter $d=\sigma_C/\sigma_{PS}=10$ and for $M=80$, 100, 120, 200 and 500 from top to bottom,
respectively. The dotted line marks the line of critical points of the demixing transition
for different chain lengths.}
\end{figure}

Figure \ref{fig:1} shows the fluid-fluid phase equilibria (solid lines), the spinodals
(dashed lines) and the critical points (black dots) for model athermal mixtures
of colloids and excluded-volume polymers considered in this work.
The calculations were carried out for the constant colloid to polymer segment 
size ratio $d=\sigma_C/\sigma_{PS}=10$ and for $M=80$, 100, 120, 200 and 500, from top to bottom, respectively. 
The diagrams are plotted in the colloid packing fraction -- reduced pressure representation, which
will be useful in the presentation of the numerical results described below.
Note that the reciprocal of the reduced pressure plays the role similar to the temperature in the case of
simple fluids, consequently the tie lines connecting the coexisting fluid states are horizontal.
The Wertheim TPT1 theory leads to the incorrect Flory exponent ($\nu=0.5$ rather than $0.588$) 
therefore the comparison in terms of the gyration radius to colloid radius size ratio $q=R_g/R_C$ is not entirely
meaningful but on the other hand, $R_g$ does not enter explicitly the numerical calculations at any point.
Therefore, assuming that the polymer gyration radius scales as
$R_g=a_P M^{\nu}\sigma_{PS}$ with $a_P\approx0.5$ \cite{Dautenhahn94,Lue00,Vega00},
our systems would correspond to $q=0.894$, 1, 1.095, 1.414, and 2.236
for $M=80$, 100, 120, 200, and 500, respectively.
Thus we consider both regimes $q<1$ and $q>1$.
We note that the reduced critical pressure $\beta P^{(cr)}\sigma_C^3$ and 
the critical colloid packing fraction $\eta_C^{(cr)}$
decrease as the chain length increases. However, as mentioned in Ref.~\cite{Paricaud03},
for $M\to\infty$ and for fixed $d$ both $\beta P^{(cr)}_{\infty}\sigma_C^3$ and $\eta^{(cr)}_{C,\infty}$ tend to a finite,
nonzero value. For $d=10$ we find that $\beta P^{(cr)}_{\infty}\sigma_C^3=0.725456$ and $\eta^{(cr)}_{C,\infty}=0.176477$
with the leading order correction $\sim {\cal O}(1/\sqrt M)$.
This result is in a qualitative agreement with the phase behavior in the so-called ``protein limit''
of colloid-polymer mixtures \cite{Sear02,Bolhuis03,Bryk03b}.

\begin{figure}
\includegraphics[clip,width=6.5cm]{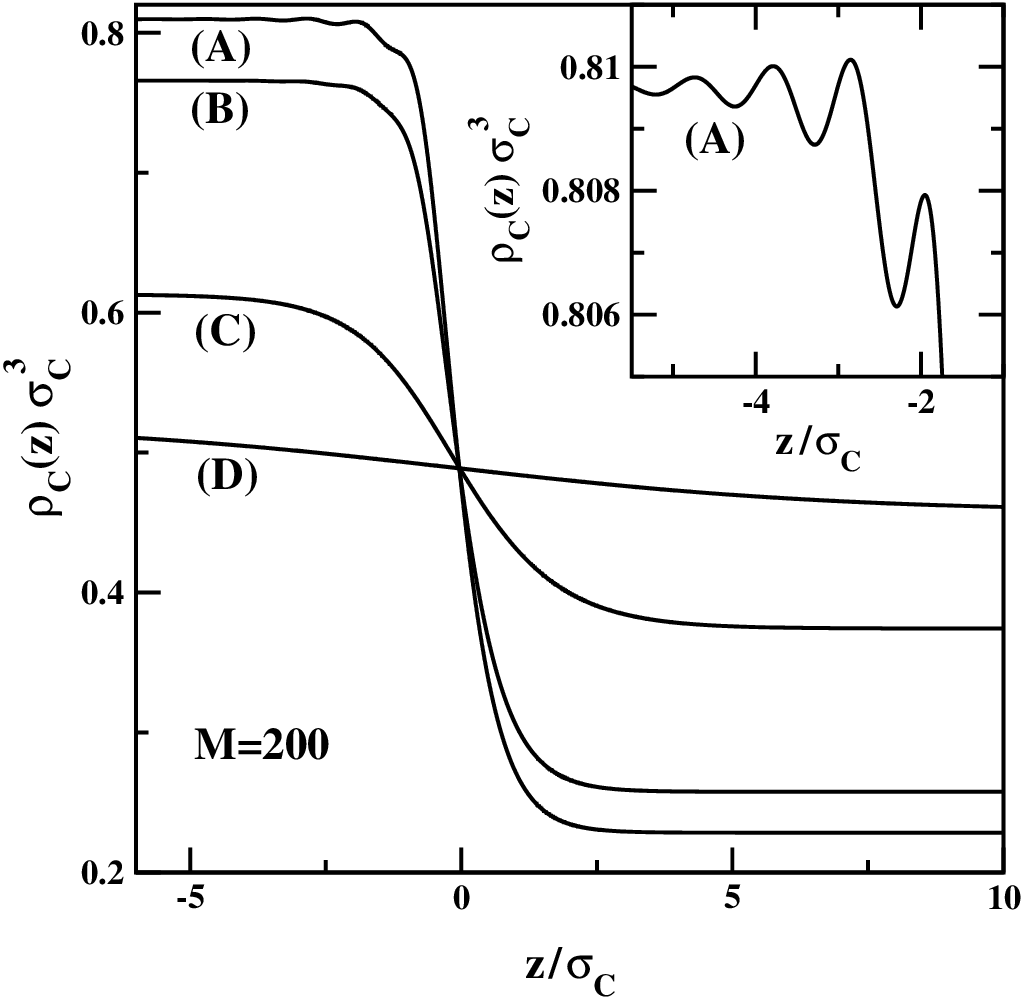}
\caption{\label{fig:2}
Colloid density profiles across the liquid-liquid interface calculated for the system
with $d=10$ and $M=200$. The profiles are evaluated for the reduced pressures
$\beta P\sigma_C^3=7$ (A), 6 (B), 4 (C) and 3.58 (D). The inset shows an enlargement of the profile
}
\end{figure}

\begin{figure}
\includegraphics[clip,width=6.5cm]{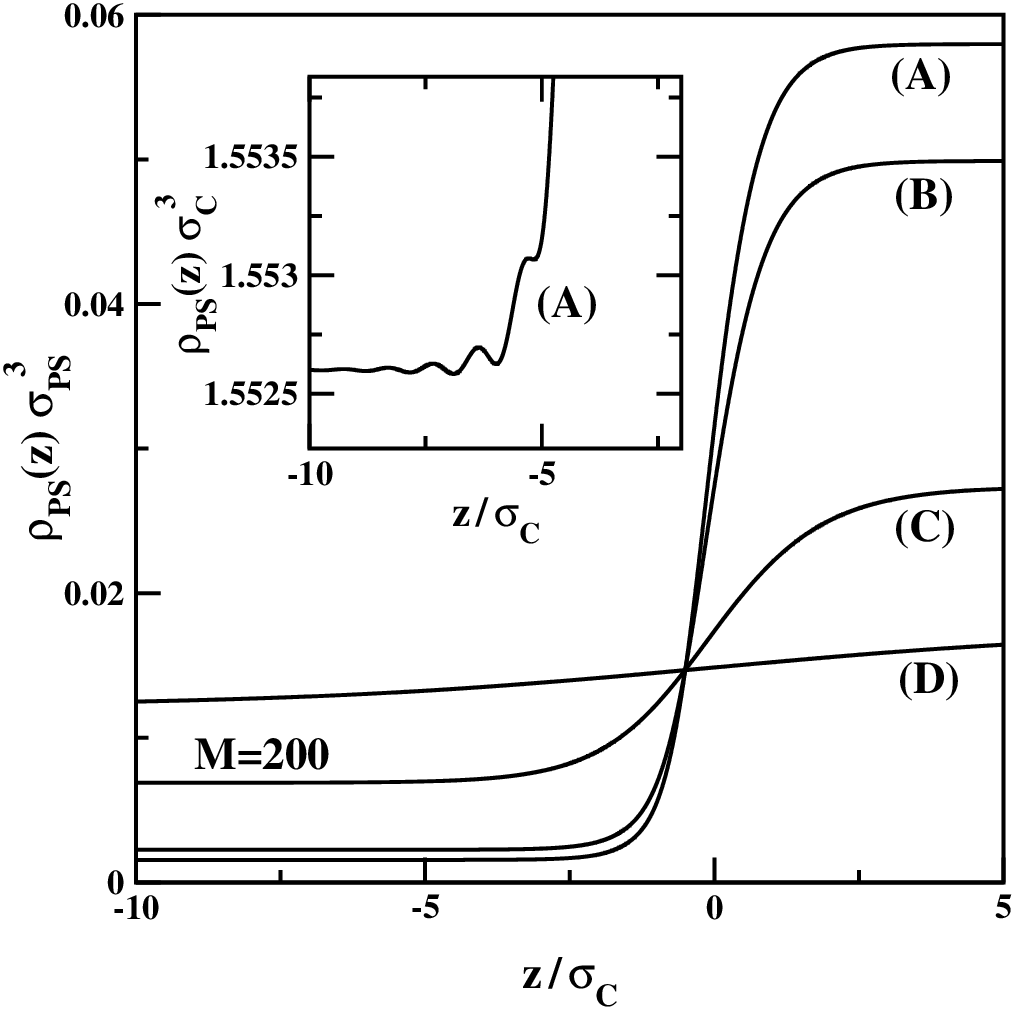}
\caption{\label{fig:3}
Average polymer segment density profiles across the liquid-liquid interface calculated for the system
with $d=10$ and $M=200$. The profiles are evaluated for the reduced pressures
$\beta P\sigma_C^3=7$ (A), 6 (B), 4 (C) and 3.58 (D).}
\end{figure}

We turn now to the interface between the demixed fluid phases.
In Figs.~\ref{fig:2} and \ref{fig:3} we display the representative examples of the colloid
and average polymer segment density profiles
evaluated for $M=200$ and for the reduced pressures $\beta P\sigma_C^3=7$ (A), 6 (B), 4 (C) and 3.58 (D).
The profiles close to the critical point are diffuse and smooth. However for larger pressures
oscillations develop on the colloid-rich side of the profiles (see the insets). It is interesting to note that
the consecutive maxima of these oscillations are located at distances approximately equal the colloid diameter.
The presence of the oscillations on both profiles can be explained by invoking the theory of the
asymptotic decay of the correlation functions in simple fluids \cite{Evans94}.
However, here the polymer is represented as a chain consisting of 200 monomers and it is not clear
how the arguments of Ref.~\cite{Evans94} can be extended to complex fluids. Nevertheless it is good to see
that the results from the present work are qualitatively similar to the findings of Brader {\it et al.} 
\cite{Brader02a}.

\begin{figure}
\includegraphics[clip,width=6.5cm]{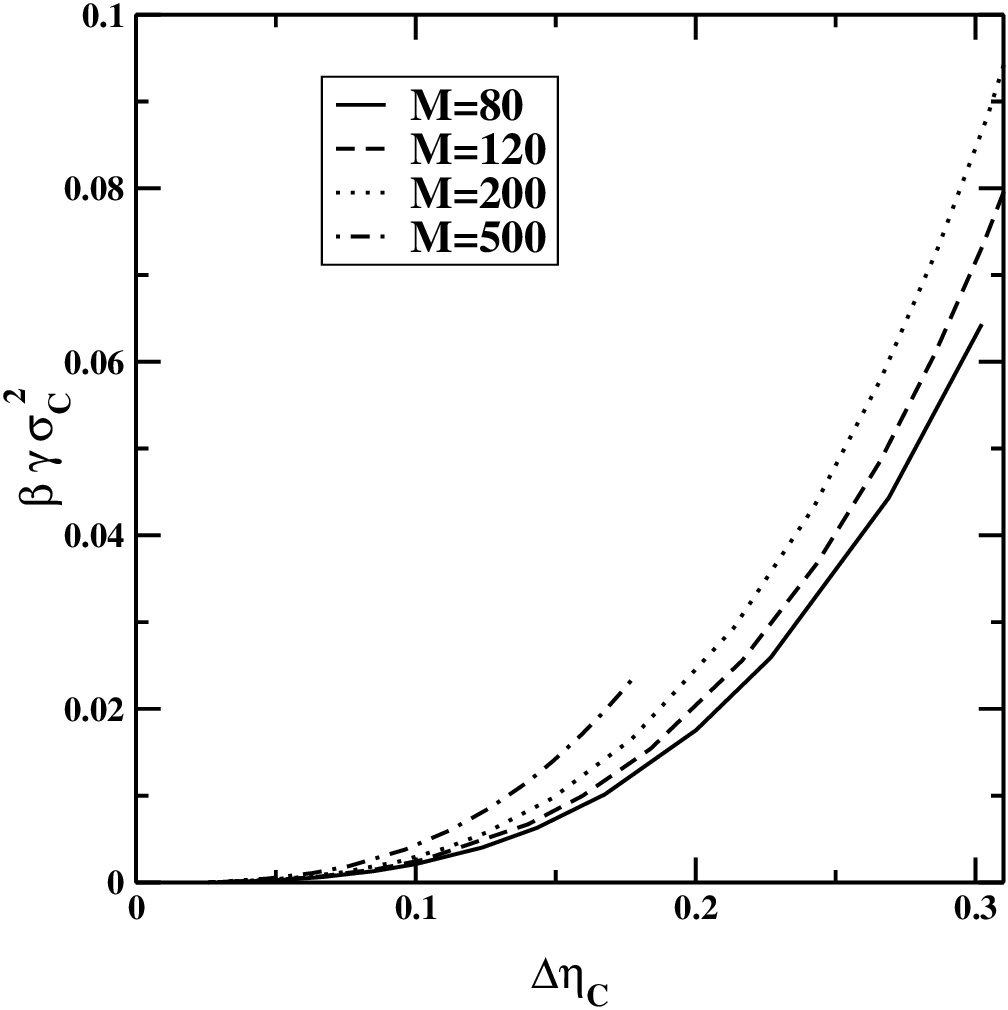}
\caption{\label{fig:4}
The reduced surface tension $\beta\gamma\sigma_C^2$ plotted against the difference in the colloid
packing fractions in the coexisting liquid (colloid-rich, L) and vapor (colloid-poor, V) phases,
$\Delta\eta_C=\eta_C^{(L)}-\eta_C^{(V)}$. The chain lengths are given in the Figure.}
\end{figure}

Figure \ref{fig:4} shows the reduced interfacial tension  $\beta\gamma\sigma_C^2$ as a function of 
the order parameter $\Delta\eta_C$ i.e. the difference in the colloid packing fractions
in coexisting ``liquid'' (colloid-rich) and vapor (colloid-poor) phases evaluated
for the chain lengths $M=80$ (solid line), 120 (dashed line), 200 (dotted line) and 500 (dot-dashed line). 
We find that the surface tension increases with the chain length. Note that the reduced surface tension
is significantly smaller than that of ideal polymers. These findings are in accordance with the recent results obtained from the squar gradient approximation
\cite{Aarts04b,Moncho03b}.

\begin{figure}
\includegraphics[clip,width=6.5cm]{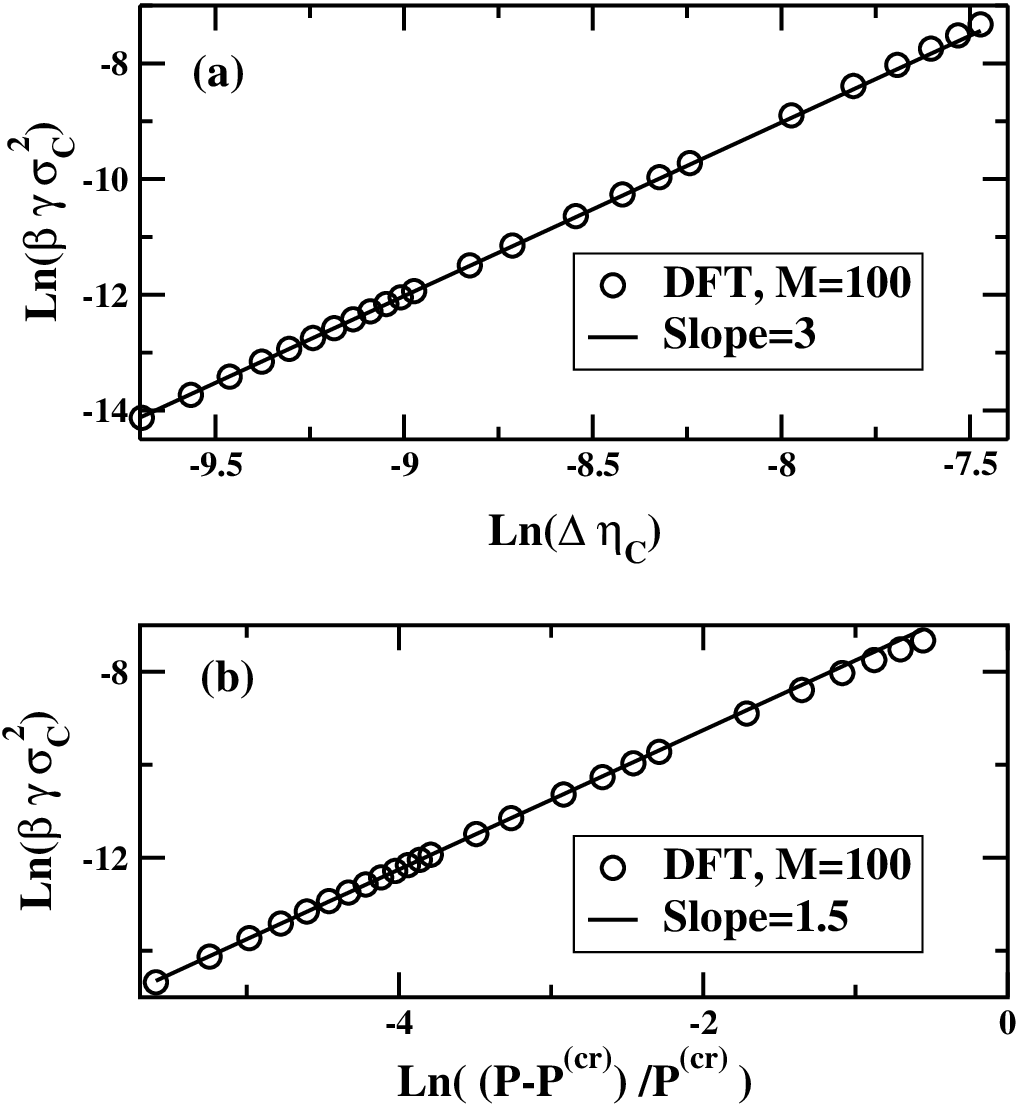}
\caption{\label{fig:5}
Critical behavior of the reduced surface tension of colloids with $d=10$
and excluded-volume polymers with $M=100$.
The open circles denote the DFT results whereas the solid lines have slopes corresponding
to the classical (mean-field) critical exponents.
(a) The logarithm of the reduced surface tension $\beta\gamma\sigma_C^2$ plotted against
the logarithm of the difference in the colloid packing fractions in the coexisting liquid
(colloid-rich, L) and vapor (colloid-poor, V) phases.
(b) The logarithm of the reduced surface tension $\beta\gamma\sigma_C^2$ plotted as a function of
the logarithm of the  relative distance to the critical point in terms of the
pressure difference, $(P-P^{(cr)})/P^{(cr)}$.
}
\end{figure}

In Fig.~\ref{fig:5} we examine the critical behavior of the interfacial tension resulting
from the present theory. Part a shows the log-log plot of the interfacial tension vs. the order parameter
$\Delta\eta_C$ whereas part b displays the logarithm of the interfacial tension plotted against
the logarithm of the relative distance to the critical point measured in terms of $(P-P^{(cr)})/P^{(cr)}$
(i.e. the analog of the reduced temperature $\tau=(T^{(cr)}-T)/T^{(cr)}$ for simple fluids).
We find that on approaching the critical point the surface tension vanishes with classical (mean field)
exponents. This is not surprising in view of the fact that the Wertheim TPT1 theory can be derived
by considering the chemical potential of a single chain in the reference solvent of monomers
\cite{Paricaud03,BenNaim71}.
Similar results have been obtained in a related study of binary polymer blends \cite{Bryk04}.

\begin{figure}
\includegraphics[clip,width=6.5cm]{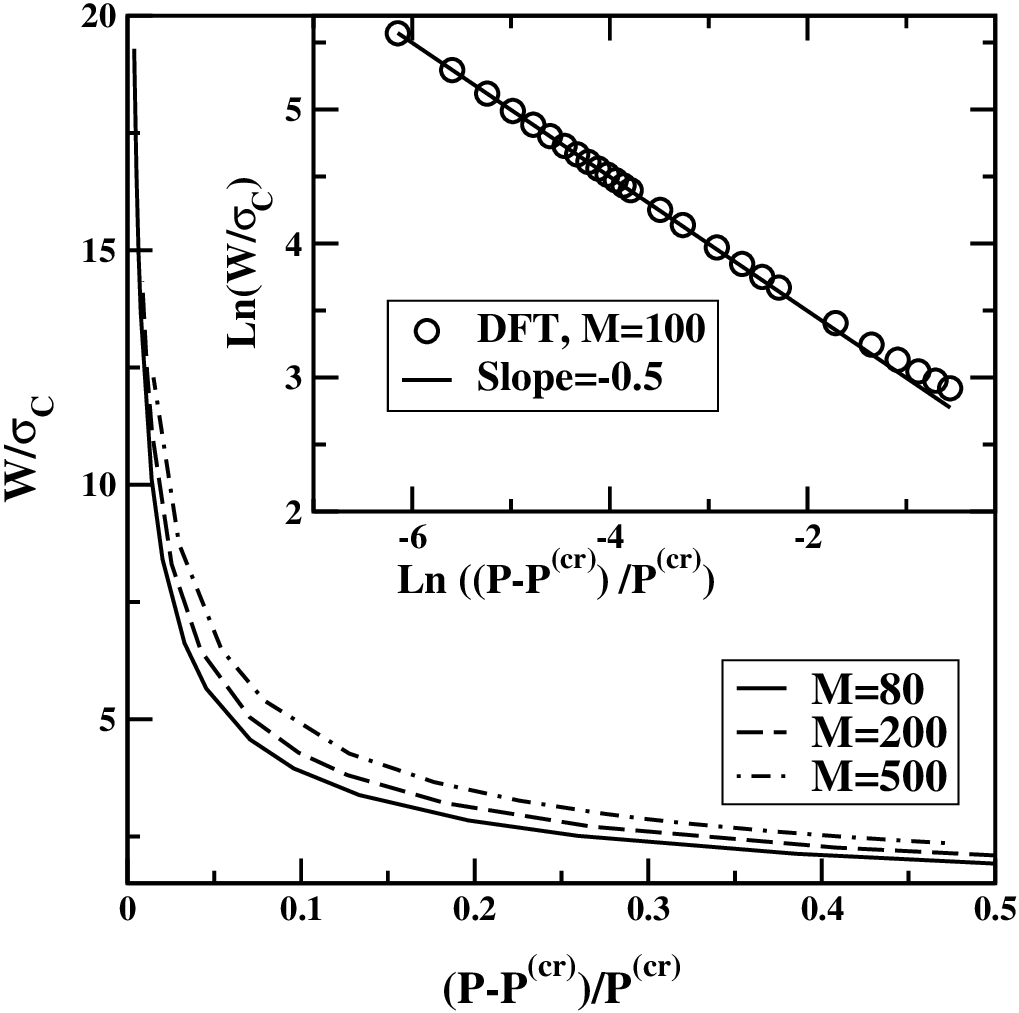}
\caption{\label{fig:6}
The interfacial width $W$ as a function of the relative distance to the critical
point expressed in terms of the pressure difference $(P-P^{(cr)})/P^{(cr)}$.
The main plot shows the interfacial width calculated for $M=80$ (solid line), 200 (dashed line)
and 500 (dot-dashed line). The inset shows a double-logarithmic plot of the same quantities
evaluated for $M=100$. The open circles denote the DFT results whereas the solid lines have slopes corresponding
to the classical (mean-field) critical exponents.
}
\end{figure}

The width of the interface can be quantitatively characterized by
the parameter $W$, defined as \cite{Fischer80}
\begin{equation}\label{eq:22}
W=[\rho_{PS}(z=\infty)-\rho_{PS}(z=-\infty)]\left[{\frac{d\rho_{PS}(z)}{dz}}\right]^{-1}_{z=z_0},
\end{equation}
where $z_0$ is given by $\rho_{PS}(z_0)=(1/2)[\rho_{PS}(z=\infty)+\rho_s(z=-\infty)]$
and $\rho_{PS}(z=\infty)$ and $\rho_{PS}(z=-\infty)$ are the segment densities
of the coexisting polymer-rich and polymer-poor phases. In Fig.~\ref{fig:6} we plot
the interfacial width calculated for the systems with $M=80$ (solid line), 200 (dashed line)
and 500 (dot-dashed line). Similar to other studies \cite{Moncho03b},
we find that the interfacial width increases with the chain length.
For the states well removed from the critical point it is of the order of two colloid diameters.
On approaching the critical point $W$ diverges as expected. In simple fluids
the interfacial width should behave similar to the correlation length $\xi$ i.e. $W\sim\xi\sim\tau^{-\nu}$.
The inset illustrates that in our approach the interfacial width indeed diverges with the mean field exponent 
$\nu=1/2$, what is consistent with the surface tension results.

\begin{figure}
\includegraphics[clip,width=6.5cm]{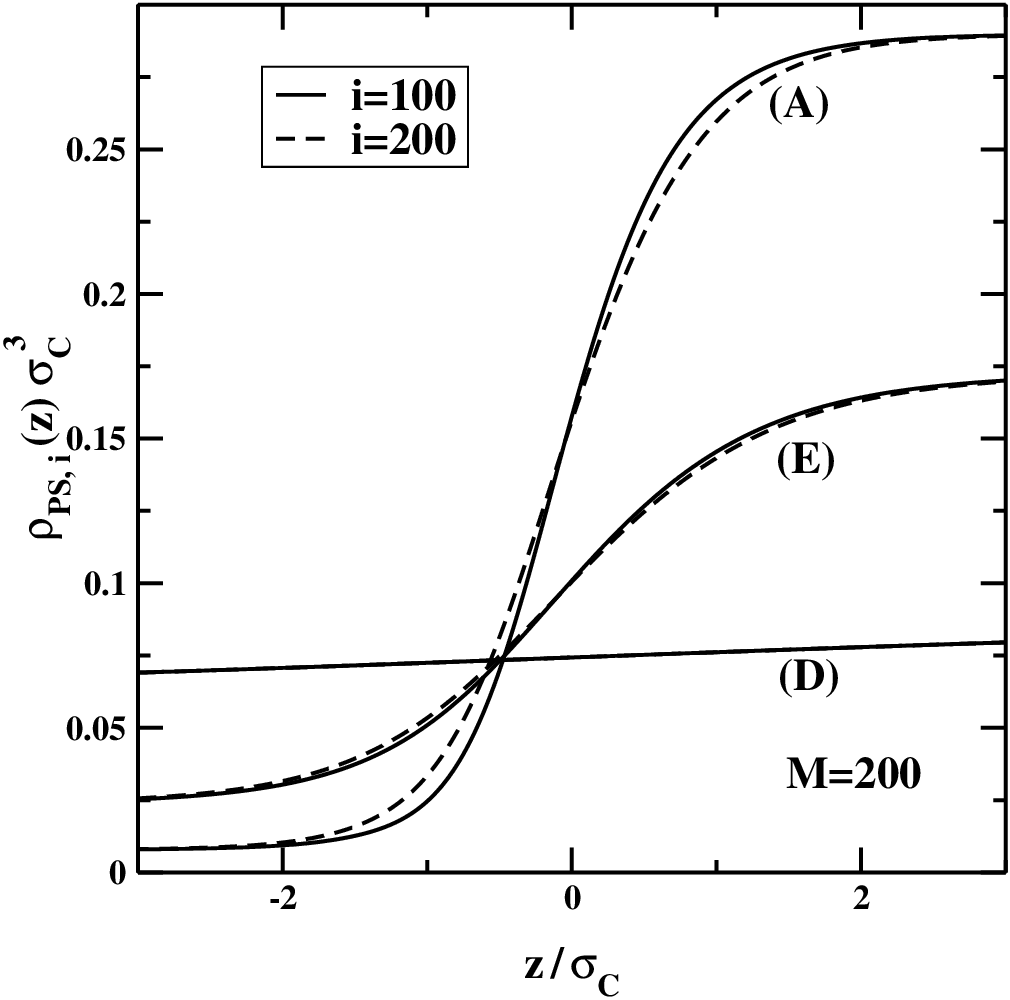}
\caption{\label{fig:7}
Middle (the solid lines) and end (the dashed lines) segment density profiles across the liquid-liquid interface
calculated for the system with $d=10$ and $M=200$.
The profiles are evaluated for the reduced pressures
$\beta P\sigma_C^3=7$ (A), 4.5 (E) and 3.58 (D).
}
\end{figure}

\begin{figure}
\includegraphics[clip,width=6.5cm]{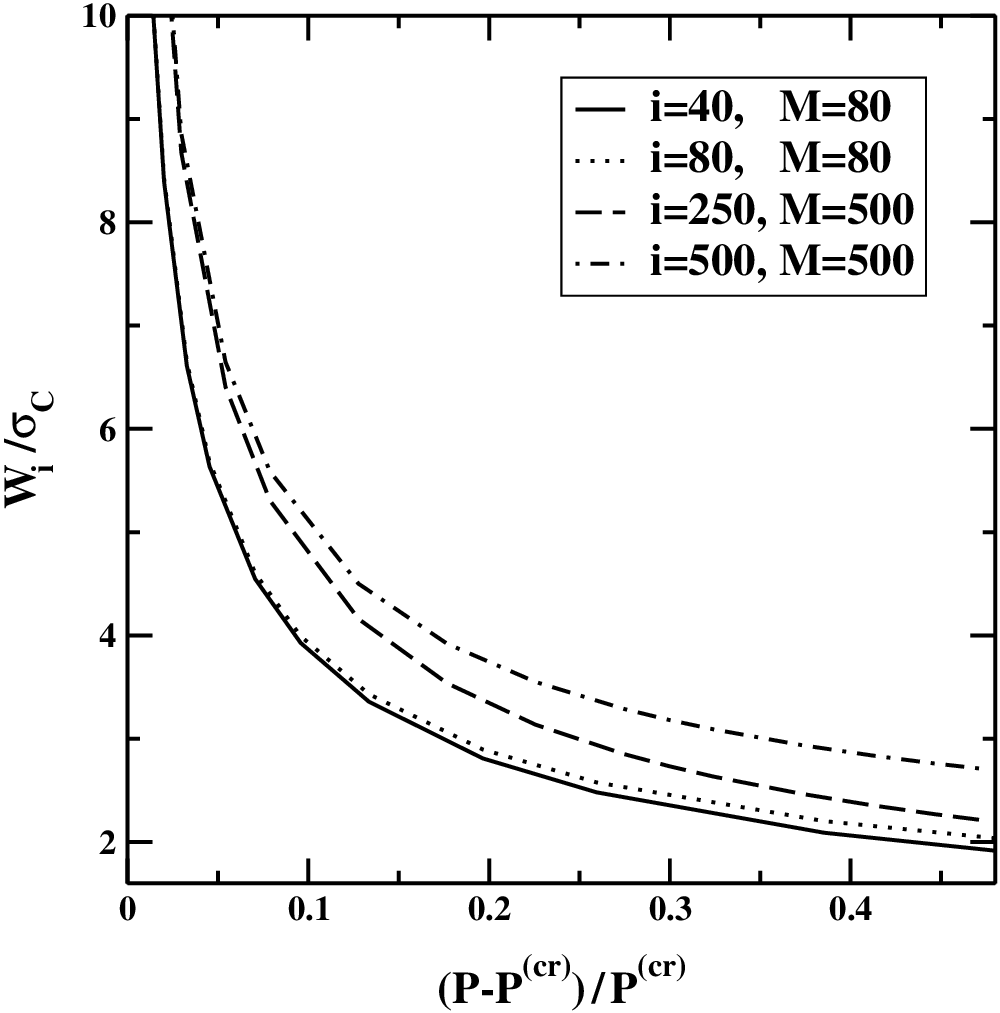}
\caption{\label{fig:8}
The interfacial width of the segment density profile $W_i$ as a function of the relative distance to the critical
point expressed in terms of the pressure difference $(P-P^{(cr)})/P^{(cr)}$.
The solid and dotted lines denote the interfacial width of the mid ($i=40$) and end ($i=80$)
segment density profile for 80-mer whereas the dashed and dot-dashed lines denote the 
interfacial width of the mid ($i=250$) and end ($i=500$) segment density profile for
500-mer.
}
\end{figure}

Finally we address the issue of the distribution of the particular segments across the interface.
Several authors have pointed out that the fluid-fluid interface in polymeric
systems is enriched in end-segments \cite{Ypma96,Kumar90,Reiter90}. It is
related to a different chain length dependence of surface
tension from bulk properties \cite{Szleifer89}.
Figure \ref{fig:7} shows the middle (solid lines) and end segment density profiles for the
systems with $M=200$ and for the reduced pressures $\beta P\sigma_c^3=7$ (A), 4.5  (E) and 3.85 (D).
For the system well removed from the critical point [marked as (A)], one observes a visible
difference between the segment profiles. The interface of the middle segments is sharper
whereas the interface of the end segments is more diffuse. For the intermediate
value of the pressure [system (E)] the difference in the distribution becomes less noticeable, and
in the vicinity of the critical point [system (D)] the distributions of the middle and end segments
are practically identical. This behavior is more pronounced for longer chains, as it is clearly
visible in Figure \ref{fig:8}, where we plot the interfacial width of the middle and end segments
evaluated for the chains with $M=80$ (solid and dotted lines) and for the longest chains considered
in this work, $M=500$ (dashed and dot-dashed lines). For states far away from the critical point
the difference of the interfacial width for $M=80$ is around one polymer segment diameter
while for $M=500$ is around the colloid radius. We conclude that for still longer chains the difference
in the distribution of the middle and end segments may be as large as colloid diameter which
in turn should be significant since the interfacial width of the average polymer segment
is of the order of 2-3 colloid diameters. This finding cannot be ascertained if the coarse-grained approach
to the description of the polymer is employed.
\section{Conclusions}
We have studied the interfacial properties of model athermal mixtures of colloids and excluded volume polymers.
The colloid particles are modeled as hard spheres while the polymer coils are modeled as freely jointed tangentially bonded
hard sphere chains. Model systems have been studied within the framework of the nonlocal density functional theory
for the fixed ratio of the colloid to polymer segment diameters and for different chain lengths.
We have found that the interfacial tension of the interacting polymer-colloid mixtures increases
with the chain length and is significantly smaller than that of the ideal polymers. These results are in accordance
with the recent studies obtained within the square gradient approach \cite{Aarts04b,Moncho03b}. For certain 
parameters we find oscillations on the colloid-rich parts of the density profiles of both colloids 
and polymers with the oscillation period of the order of the colloid diameter. The interfacial width is 
few colloid diameters wide and also increases with the chain length. We find the interfacial width
for the end segments to be larger than that for the middle segments and this effect is more pronounced
for longer chains. Within our approach the interfacial properties close to the critical point 
behave in a mean field fashion.

Before the quantitative comparison with experiments is possible
some important the issues associated with the framework proposed by Paricaud {\it et al}. \cite{Paricaud03}
should be solved. In particular, further research should focus on
tackling the problem of the inaccurate description of the polymer in the dilute limit and
the incorporation of the fluid-solid equilibria into the framework.
However, despite these deficiencies the microscopic model of the colloids-excluded volume polymer mixtures
studied in the present work seems to be an interesting alternative to the coarse-grained approaches.

\begin{acknowledgments}
This work has been supported by KBN of Poland under the Grant
1P03B03326 (years 2004-2006).
\end{acknowledgments}

\newpage

\end{document}